\documentclass{emulateapj}
\usepackage{epstopdf}
\newcommand {\aplt} {\ {\raise-.5ex\hbox{$\buildrel<\over\sim$}}\ } 
\newcommand {\apgt} {\ {\raise-.5ex\hbox{$\buildrel>\over\sim$}}\ } 

\def\sun{\hbox{$\odot$}}

\newcommand{\gps}{\ensuremath{g_{\rm P1}}}
\newcommand{\rps}{\ensuremath{r_{\rm P1}}}
\newcommand{\ips}{\ensuremath{i_{\rm P1}}}
\newcommand{\zps}{\ensuremath{z_{\rm P1}}}
\newcommand{\yps}{\ensuremath{y_{\rm P1}}}

\newcommand{\PS}{\protect \hbox {Pan-STARRS1}}

\slugcomment{Accepted for Publication in ApJ}
\shorttitle{PS1-13arp}
\shortauthors{Gezari et al.}

\begin{document}

\title{\textsl{GALEX} Detection of Shock Breakout in Type II-P Supernova PS1-13\lowercase{arp}: Implications for the Progenitor Star Wind}

\author{
S. Gezari\altaffilmark{1}, 
D. O. Jones\altaffilmark{2}, 
N. E. Sanders\altaffilmark{3},
A. M. Soderberg\altaffilmark{3},
T. Hung\altaffilmark{1},
S. Heinis\altaffilmark{1},
S. J. Smartt\altaffilmark{4},
A. Rest\altaffilmark{5}, 
D. Scolnic\altaffilmark{6}, 
R. Chornock\altaffilmark{7},
E. Berger\altaffilmark{3},
R. J. Foley\altaffilmark{8,9},
M. E. Huber\altaffilmark{10},
P. Price\altaffilmark{11}
C. W. Stubbs\altaffilmark{3,12},
A. G. Riess\altaffilmark{2},
R. P. Kirshner\altaffilmark{3,12},
K. Smith\altaffilmark{4},
W. M. Wood-Vasey\altaffilmark{13},
D. Schiminovich\altaffilmark{14},
D. C. Martin\altaffilmark{15},
W. S. Burgett\altaffilmark{16},
K. C. Chambers\altaffilmark{10},
H. Flewelling\altaffilmark{10},
N. Kaiser\altaffilmark{10},
J.L. Tonry\altaffilmark{10},
\&
R. Wainscoat\altaffilmark{10}
}
\altaffiltext{1}{Department of Astronomy, University of Maryland, Stadium Drive, College Park, MD 20742-2421, USA \email{suvi@astro.umd.edu}}
\altaffiltext{2}{Department of Physics and Astronomy, Johns Hopkins University, 3400 North Charles Street, Baltimore, Maryland 21218, USA}
\altaffiltext{3}{Harvard-Smithsonian Center for Astrophysics, 60 Garden Street, Cambridge, Massachusetts 02138, USA}
\altaffiltext{4}{Astrophysics Research Centre, School of Mathematics and Physics, Queen's University Belfast, Belfast BT7 1NN, UK}
\altaffiltext{5}{Space Telescope Science Institute, 3700 San Martin Drive, Baltimore, Maryland 21218, USA}
\altaffiltext{6}{Kavli Institute for Cosmological Physics, University of Chicago, 5640 South Ellis Avenue, Chicago, IL 60637}
\altaffiltext{7}{Ohio University, 1 Ohio University, Athens, OH 45701, USA}
\altaffiltext{8}{Astronomy Department, University of Illinois at Urbana-Champaign, 1002 West Green Street, Urbana, IL 61801, USA}
\altaffiltext{9}{Department of Physics,University of Illinois Urbana-Champaign,1110 W.\ Green Street, Urbana, IL 61801 USA}
\altaffiltext{10}{Institute for Astronomy, University of Hawaii, 2680 Woodlawn Drive, Honolulu, Hawaii 96822, USA}
\altaffiltext{11}{Department of Astrophysical Sciences, Princeton University, 4 Ivy Lane, Princeton, NJ, 08544, USA}
\altaffiltext{12}{Department of Physics, Harvard University, USA}
\altaffiltext{13}{Pittsburgh Particle Physics, Astrophysics, and Cosmology Center, Department of Physics and Astronomy, University of Pittsburgh, 3941 O'Hara Street, Pittsburgh, Pennsylvania 15260, USA}
\altaffiltext{14}{Department of Astronomy, Columbia University, 550 West 120th Street, New York, NY 10027, USA}
\altaffiltext{15}{AstronomyDepartment, California Institute of Technology, MC 249-17, 1200 East California Boulevard, Pasadena, CA 91125, USA}
\altaffiltext{16}{GMTO Corporation, 251 S. Lake Ave., Suite 300, Pasadena, CA 91101, USA}

\begin{abstract}
We present the \textsl{GALEX} detection of a UV burst at the time of explosion of an optically normal Type II-P supernova (PS1-13arp) from the Pan-STARRS1 survey at $z=0.1665$.  The temperature and luminosity of the UV burst match the theoretical predictions for shock breakout in a red supergiant, but with a duration a factor of $\sim$50 longer than expected.  We compare the $NUV$ light curve of PS1-13arp to previous \textsl{GALEX} detections of Type IIP SNe, and find clear distinctions that indicate that the UV emission is powered by shock breakout, and not by the subsequent cooling envelope emission previously detected in these systems.  We interpret the $\sim 1$d duration of the UV signal with a shock breakout in the wind of a red supergiant with a pre-explosion mass-loss rate of $\sim 10^{-3}$ $M_\odot$ yr$^{-1}$.  This mass-loss rate is enough to prolong the duration of the shock breakout signal, but not enough to produce an excess in the optical plateau light curve or narrow emission lines powered by circumstellar interaction.  This detection of non-standard, potentially episodic high mass-loss in a RSG SN progenitor has favorable consequences for the prospects of future wide-field UV surveys to detect shock breakout directly in these systems, and provide a sensitive probe of the pre-explosion conditions of SN progenitors.  \end{abstract}

\keywords{ultraviolet: general --- surveys -- supernovae}

\section{Introduction}

The first electromagnetic signature of a core-collapse SN is expected to occur when the explosion shock wave accelerates through the stellar envelope, and the radiation from the shock escapes and ``breaks out'' of the stellar surface \citep{Falk1978, Klein1978}.  
This radiative precursor to the SN explosion is expected to be the most luminous phase of a SN, and yet it is the least well observed.  This is because the associated burst of radiation is brief (1--1000 sec)  and hot ($\sim 10^5-10^6$ K) \citep{Blinnikov2002, Ensman1992, Klein1978}, and {\it precedes} the optically-bright phase during which SNe are typically discovered.  

These obstacles to discovery were surmounted with a bit of luck in a pointed X-ray observation of a galaxy in which a Type Ibc SN fortuitously exploded, and a several hundred second X-ray flash was detected \citep{Soderberg2008, Modjaz2009}, with a duration and spectral shape that was consistent with shock breakout into the thick wind of a Wolf-Rayet (WR) star progenitor \citep{Balberg2011, Svirski2014}.  The presence of a thick wind is expected for a WR star, but it was also an important factor in stretching the shock breakout timescale long enough (a factor of 100 longer than expected for shock breakout through the stellar surface) to be easily detectable.  In fact, there have been reports of shock breakout detections in Type II SNe on the timescale of hours to several days \citep{Schawinski2008, Ofek2010, Ofek2013, Drout2014}, which require an unusual amount of circumstellar material in order to explain the timing of the events, either in the form of an extended stellar envelope inconsistent with stellar evolution models \citep{Schawinski2008}, or a vigorous, episodic mass-loss rate of $\approx 0.1 M_\odot$ yr$^{-1}$ \citep{Ofek2010, Drout2014}.

\begin{figure*}[htp]
\begin{center}
\includegraphics[scale=0.5]{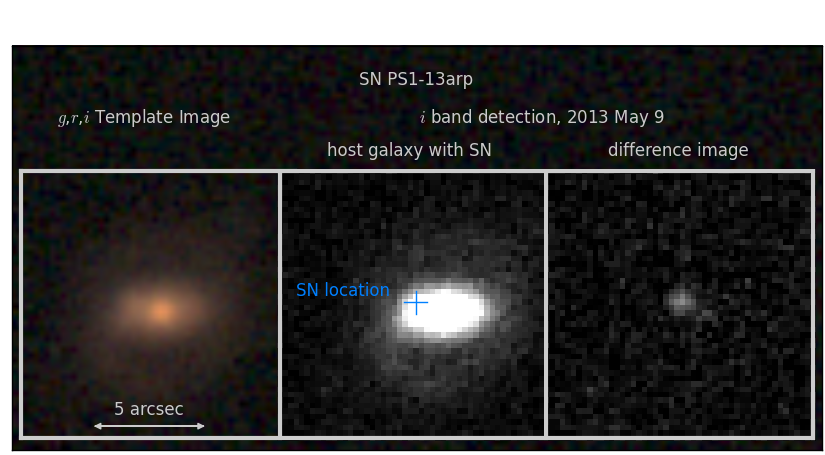}
\caption{PS1 deep $\gps, \rps, \ips$\ reference template ({\it left}), $i$-band image on UT 2013 May 9 ({\it middle}), and $i$-band difference image ({\it right}).  The position of PS1-13arp in the host galaxy is marked with a blue cross.   \label{fig:diff}}

\end{center}
\end{figure*}

A few of the earliest optical discoveries of core-collapse SNe have caught the decline of the ``cooling envelope phase'' following shock breakout \citep{Hamuy1988, Stritzinger2002, Dessart2008}.  Observing the {\it rise} to peak of the cooling envelope emission is more desirable because it puts direct constraints on the radius of the progenitor star \citep{Nakar2010, Rabinak2011}.   This thermal emission from the expanding shock-heated ejecta lasts only minutes in the soft X-rays,  but peaks in the UV at hours up to a couple days after shock breakout \citep{Nakar2010}.  By the time the thermal emission peaks in the optical, other radiative processes dominate, i.e. radioactivity (and hydrogen recombination in Type II SNe).   The timescales of the UV peak are more accessible to SN surveys, however, they still require observing the SN in the UV {\it before} it brightens in the optical, which can be days later.  With \textsl{GALEX} we have attempted to circumvent this problem by coordinating wide-field monitoring in the UV with ground-based monitoring in the optical.  

The large radii of red supergiant (RSG) stars yield longer timescales and larger luminosities during shock breakout and its immediate aftermath, making them appealing for observational searches.  Stellar evolution calculations predict that RSG stars will explode as Type IIP SNe \citep{Heger2003}, and Type IIP SNe have been directly linked with RSG progenitors from pre-explosion imaging \citep{Smartt2009}.  Indeed, the first early UV detections of SNe were of two Type IIP SNe, which were discovered serendipitously, from overlap between the \textsl{GALEX} Deep Imaging Survey and the CFHT Supernova Legacy Survey \citep{Gezari2008, Schawinski2008}.  

The first planned UV and optical joint survey effort, the \textsl{GALEX} Time Domain Survey \citep{Gezari2013} in the $NUV$, was coordinated with the optical Pan-STARRS1 Medium Deep Survey (PS1 MDS), and yielded the early detection of the rise to peak of UV emission from a Type IIP SN at $z=0.0862$ with a cadence of 2 days.  The early UV peak was well fitted by cooling envelope emission from the explosion of a red supergiant (RSG) with $R = 700 \pm 200 R_\odot$ \citep{Gezari2010}.  While the \textsl{GALEX} observations caught the rise to the peak of the cooling envelope emission, the shock breakout itself was not resolved by the relatively coarse time sampling of the observations compared to the expected shock breakout  duration ($< 1$hr).   \citet{Ganot2014} present a shallower, wide-field joint $NUV$ \textsl{GALEX} and optical Palomar Transient Factory (PTF) survey which detected 7 core-collapse SNe in the UV with a 3 day cadence.  Their observed detection rate is consistent with core-collapse SN rates combined with a fiducial model for cooling envelope emission from RSG stars.  However, again their cadence was not sufficient to catch the short-lived shock breakout signal.  

Here we present the results from a high-cadence monitoring program with \textsl{GALEX} which increased the time sampling of the \textsl{GALEX} Time Domain Survey and GALEX/PTF survey by a factor of two and three, respectively, and has yielded the first potential detection of the elusive shock breakout phase in a Type IIP SN, albeit with the requirement that the shock breakout occurred outside the progenitor's surface, in its circumstellar wind.  The paper is organized as follows, in \S \ref{sec:obs} we describe the \textsl{GALEX} and PS1 MDS observations and Gemini spectroscopy, in \S \ref{sec:disc} we compare the UV/optical light curve of PS1-13arp to previous early observations of Type IIP SNe, in \S \ref{sec:models} we compare the observed UV burst to models for shock breakout from the surface of a star, cooling envelope emission, and shock breakout into a wind, and in \S \ref{sec:conc} we conclude with our favored model of shock breakout of an RSG with a pre-explosion mass-loss rate of $\approx 10^{-3} M_\odot$ yr$^{-1}$.
 
\section{Observations} \label{sec:obs}

\subsection{Pan-STARRS1 Medium Deep Survey}

The Pan-STARRS1 system is a high-etendue wide-field imaging system, designed for dedicated survey observations. The system is installed on the peak of Haleakala on the island of Maui in the Hawaiian island chain. 
A description of the Pan-STARRS1 system, both hardware and software, is provided by \cite{PS1_system}.
The Pan-STARRS1 optical design \citep{PS1optics} uses a 1.8~meter diameter $f$/4.4 primary mirror, and a 0.9~m secondary, and a 3.3 deg diameter field of view. 
The Pan-STARRS1 imager \citep{PS1_GPCA} comprises a total of 60 $4800\times4800$ pixel detectors, with 10~$\mu$m pixels that subtend 0.258~arcsec.  

%note trailing slash command where space required after passband macros
The PS1 observations obtained through a set of five broadband
filters, which we have designated as \gps, \rps, \ips, \zps, and \yps.   Although the filter system for \PS\ has much in
common with that used in previous surveys, such as SDSS \citep{SDSS}, there
are important differences.  Further information on the passband shapes is described
in \cite{PS_lasercal}.   Photometry is in the ``natural'' \PS\  
system, $m=-2.5\log(flux)+m'$, with a single zeropoint adjustment $m'$ made in each band to conform to the AB magnitude scale \citep{JTphoto}. 

\begin{figure*}[htp]
\begin{center}
\includegraphics[scale=0.75]{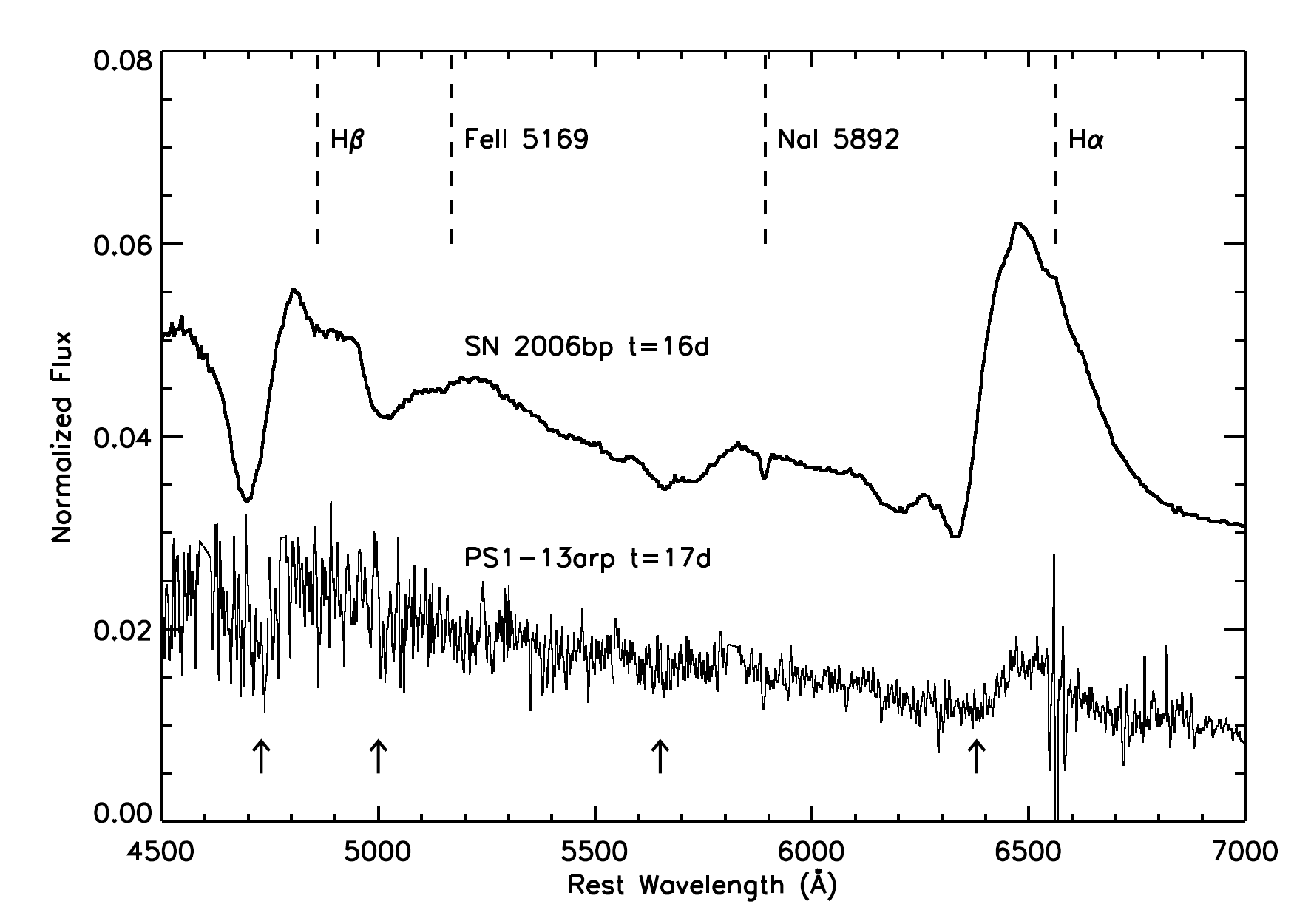}
\caption{Gemini spectrum of PS1-13arp (smoothed by 2 pixels) at $\sim$ 17 rest-frame days since shock breakout, in comparison to SN 2006bp at 16 rest-frame days since shock breakout.  Arrows show the potential P-Cygni absorption features associated with broad $H\beta$, FeII $\lambda 5169$, NaI $\lambda 5892$, and H$\alpha$, whose central wavelength are marked with dashed lines.
\label{fig:fullspec}
}
\end{center}
\end{figure*}

\begin{figure}[htp]
\begin{center}
\includegraphics[scale=0.5]{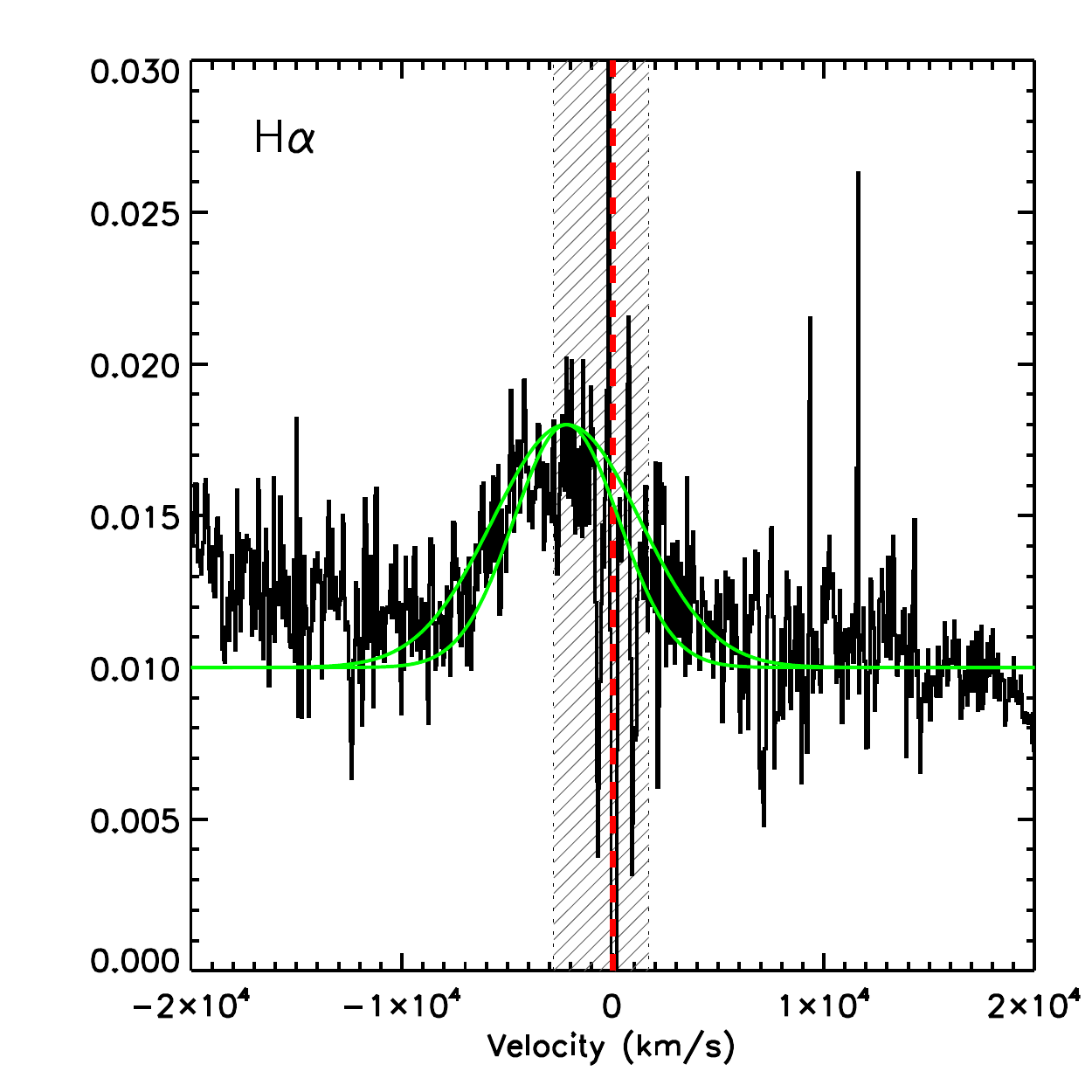}
\caption{H$\alpha$ profile of PS1-13arp at $\sim$17 rest-frame days since shock breakout.  The line is broad (FWHM = 7500 $\pm$ 1200 km/s, Gaussian plotted with green lines) and blueshifted by 2200 $\pm$ 100 km/s.  Red dashed line indicates zero velocity defined by the host galaxy H$\alpha$ line.  Hashed region indicates the location of the telluric A band feature that has been corrected for in the line profile.  The profile is mostly symmetric, and does not yet demonstrate the strong P-Cygni absorption that develops in Type IIP SNe typically 3 weeks after explosion.  
\label{fig:halpha}
}
\end{center}
\end{figure}

Images obtained by the Pan-STARRS1 system are processed through the Image Processing Pipeline (IPP) \citep{PS1_IPP}, 
on a computer cluster at the Maui High Performance Computer Center. The pipeline runs the images through a succession of stages, 
including flat-fielding (``de-trending''), a flux-conserving warping to a sky-based image plane (with a plate scale of 0.25 arcsec/pixel), masking and artifact removal, and object detection and 
photometry.  Mask and variance arrays
are carried forward at each stage of the IPP processing. Photometric and astrometric measurements performed by the IPP system are described in \cite{PS1_photometry}
and \cite{PS1_astrometry} respectively.

PS1 has two extragalactic survey modes:  the 3$\pi$ survey observes the entire sky above a declination of $-30$ deg in 5 bands with a sparse cadence (a few epochs per month per filter), and the Medium Deep Survey (MDS) is a deeper survey of 10 fields distributed across the sky in 5 filters with a daily cadence.  This paper uses images and photometry from the PS1 MDS.  
Observations of $3-5$ MD fields are taken each night and the filters are cycled through in the pattern:
\gps\ and \rps\
in the same night (dark time), followed by \ips\ and \zps\ on the subsequent second and third night, respectively.
Around full moon only \yps\ data are taken.  A nightly stack image is created using a variance-weighted scheme from 8 dithered exposures, allowing for the removal of defects like cosmic rays and satellite streaks, and resulting in a 5$\sigma$ depth of $\sim 23$ mag in the $\gps, \rps, \ips$, \zps\ bands, and $\sim 22$ mag in the \yps\ band in the AB system.  

The nightly MD stacked images are processed through a frame-subtraction analysis using the {\it photpipe} pipeline \citep{Rest2005}, and significant excursions in the difference images are detected using a modified version of DoPHOT \citep{Schechter1993}.  Reference templates are constructed for the MD fields from two epochs obtained at the beginning of the observing season.
Spectroscopically confirmed supernovae are later reprocessed using deep template images, constructed from all nightly stacks with good seeing in the seasons that the SN did not explode, with the routine {\it transphot} \citep{Rest2014} which implements the HOTPANTS image differencing algorithm \citep{Becker2004}, then PSF photometry is applied using DAOPHOT \citep{Stetson1987}.  The weighted average position of the SN is calculated, and forced photometry is then applied in every difference image at that position.  The PS1 MDS supernova search has yielded hundreds of spectroscopically confirmed SNe, including Type Ia SNe for cosmology studies \citep{Rest2014}, core-collapse SNe, including 76 SNe IIP \citep{Sanders2014}, as well as several rare, superluminous SNe at high redshift \citep{Chomiuk2011,Berger2012,Lunnan2013}.

\subsection{GALEX High-Cadence Time Domain Survey}

We conducted an extended survey with \textsl{GALEX} in the $NUV$ during December 2012 through April 2013 of 4 PS1 MD fields, in order to perform a high-cadence (1 day) time domain survey, and to complete \textsl{GALEX} coverage started with the \textsl{GALEX} Time Domain Survey \citep{Gezari2013} for 9 out of the 10 PS1 MD fields.  The $FUV$ detector on \textsl{GALEX} was no longer operational after May 2009.  The 1-day cadence of the \textsl{GALEX} observations was chosen to better probe the cooling envelope emission of RSG stars, which has a characteristic timescale of 2 days \citep{Nakar2010, Rabinak2011}, and was just barely resolved by the 2-day cadence of the \textsl{GALEX} Time Domain Survey for the Type IIP SN 2010aq \citep{Gezari2010}.  The exposure times for each \textsl{GALEX} visit averaged 1.1 ks, resulting in a 3$\sigma$ depth of 23.5 mag in the $NUV$ in the AB system.   
Here we present the joint \textsl{GALEX} and PS1 discovery of Type IIP SN PS1-13arp, for which we catch the onset of a luminous UV burst with \textsl{GALEX} that precedes its standard optical behavior observed by PS1.

\subsection{Imaging and Spectroscopy of PS1-13arp}

PS1-13arp was identified as an optical transient in PS1 field MD06 by {\it photpipe} on UT 2013 April 26 at R.A. 12:18:25.108, Dec.~+46:37:01.30 (J2000), with a spatial offset of 1.577 arcsec East and 0.275 arcsec North of its host galaxy (SDSS J121824.98+463700.5).  Figure \ref{fig:diff} shows the PS1 difference image detection in the $i$ band, as well as a color image of the deep reference template with only the host galaxy flux.  The SN was also detected as a variable source by \textsl{GALEX} during monitoring of the field PS\_NGC4258\_MOS107 at the 8.3$\sigma$ level on UT 2013 April 15.  The host galaxy is detected in the $NUV$, and measured from the 4 epochs before the SN is detected by the \textsl{GALEX} pipeline to have $NUV= 21.20 \pm 0.05$ mag, measured from a 6 arcsec radius circular aperture, and including a correction for the total energy enclosed of $-0.23$ mag \citep{Morrissey2007}.   Table \ref{tab1} gives the SN photometry from \textsl{GALEX} and PS1.

We obtained spectroscopy of the UV/optical transient with Gemini/GMOS on UT 2013 May 4.  We obtained 3 exposures with a 1.0 arcsec slit and the R400 grating of 1200 sec each, resulting in a spectral resolution of $\sim 4$ \AA, with a dither along the slit, and a change in grating central wavelength, in order to avoid bad columns, and to cover the gap in wavelength coverage for the grating.  The spectrum was overscan-corrected, bias subtracted, flat-fielded, wavelength calibrated, and flux calibrated using standard IRAF routines.  Observations of a standard star (Feige 34) were used for flux calibration and corrections for telluric absorption.   The 2D spectrum displays strong, spatially extended emission lines (H$\alpha$, [N II] $\lambda\lambda 6583, 6548$, [S II]$\lambda\lambda 6716,6731$) indicative of star-formation in the host galaxy, from which we measure a redshift of $z=0.1665 \pm 0.001$.  
 
 We fitted the rest-frame host galaxy broad-band spectral energy distribution from \textsl{GALEX} $FUV$ and $NUV$ \citep{Martin2005}, PS1 \gps, \rps, \ips, \zps, \yps\ deep stacks with $u$-band coverage from CFHT (Heinis et al., in prep.), and infrared from WISE at 3.4$\mu$m and 4.6$\mu$m \citep{Wright2010}, with the code CIGALE (Noll et al 2009), using version v0.3, implemented in Python. CIGALE combines a UV-optical stellar SED with a dust component emitting in the infrared, and ensures full energy balance between the dust absorbed stellar emission and its re-emission in the IR.
%We adopt here a model with two stellar populations: an exponentially declining SFR, and a recent stellar population with a constant SFR.
%We use here Bruzual & Charlot (2003) single stellar population synthesis models, and a Chabrier (2003) Initial Mass Function. We model the attenuation by dust with the attenuation law from Calzetti et al. (2000). The re-emission in the IR of the stellar emission absorbed by dust is modeled using the empirical templates of Dale et al. (2014).  
The fit requires both an old stellar population with an exponentially declining star formation rate with $\tau = 6.7 \pm 2$ Gyr as well as a recent stellar population with a moderate star formation rate of $3.6 \pm 1.2 M_{\odot}$ yr$^{-1}$, with a total stellar mass of $\log(M/M_\odot) = 10.4 \pm 0.1$.  Note that this mass is consistent with the gas-phase metallicity that we estimate from the ratio of the strong nebular emission lines in the host galaxy spectrum.  We measure $\log ([N II] \lambda 6584/H\alpha) = -0.37 \pm 0.06$, which corresponds to a metallicity of $\log(O/H) + 12 \approx 9$ \citep{Kewley2002}, and is in good agreement with the mass-metallicity relation for star-forming galaxies \citep{Tremonti2004}.

We extract a 1D SN spectrum by first extracting a 1D spectrum using a 5-pixel aperture centered on the SN (including SN and galaxy light), and then extract a 1D spectrum with an aperture centered on a region 5 pixels away from the SN center, with only galaxy light.  We isolate the SN spectrum by subtracting the 1D galaxy spectrum from the 1D SN+galaxy spectrum.  We coadded the first two exposures, which had the highest SN signal.  We show the coadded SN spectrum in Figure \ref{fig:fullspec}, in comparison to the spectrum of SN 2006bp at a similar phase since shock breakout from \citet{Quimby2007}.   We detect the P-Cygni absorption features of H$\beta$, Fe II$\lambda 5169$, NaI $\lambda 5892$, and H$\alpha$.  We fit the strong broad H$\alpha$ line with a Gaussian with a FWHM = $7500 \pm 1200$ km/s (shown in Figure \ref{fig:halpha}).  The line appears mostly symmetric, and blue-shifted by 2200 $\pm 100$ km/s from the galaxy's narrow H$\alpha$ line.  We only see weak evidence of blue-shifted absorption, as one would expect in a P-Cygni profile.  The weak blueward absorption is consistent with a young Type IIP SN, which typically do not develop a strong P-Cygni Balmer line shape until $\sim 3$ weeks after explosion \citep{Quimby2007, Dessart2008}.  However, the weak P-Cygni absorption could also reveal unusual conditions in the SN envelope or circumstellar environment.  Unfortunately, without a later epoch of spectroscopy, we can only speculate as to whether PS1-13arp would have developed a stronger P-Cygni line profiles at late times.

\begin{figure*}
\centering
\includegraphics[scale=0.75]{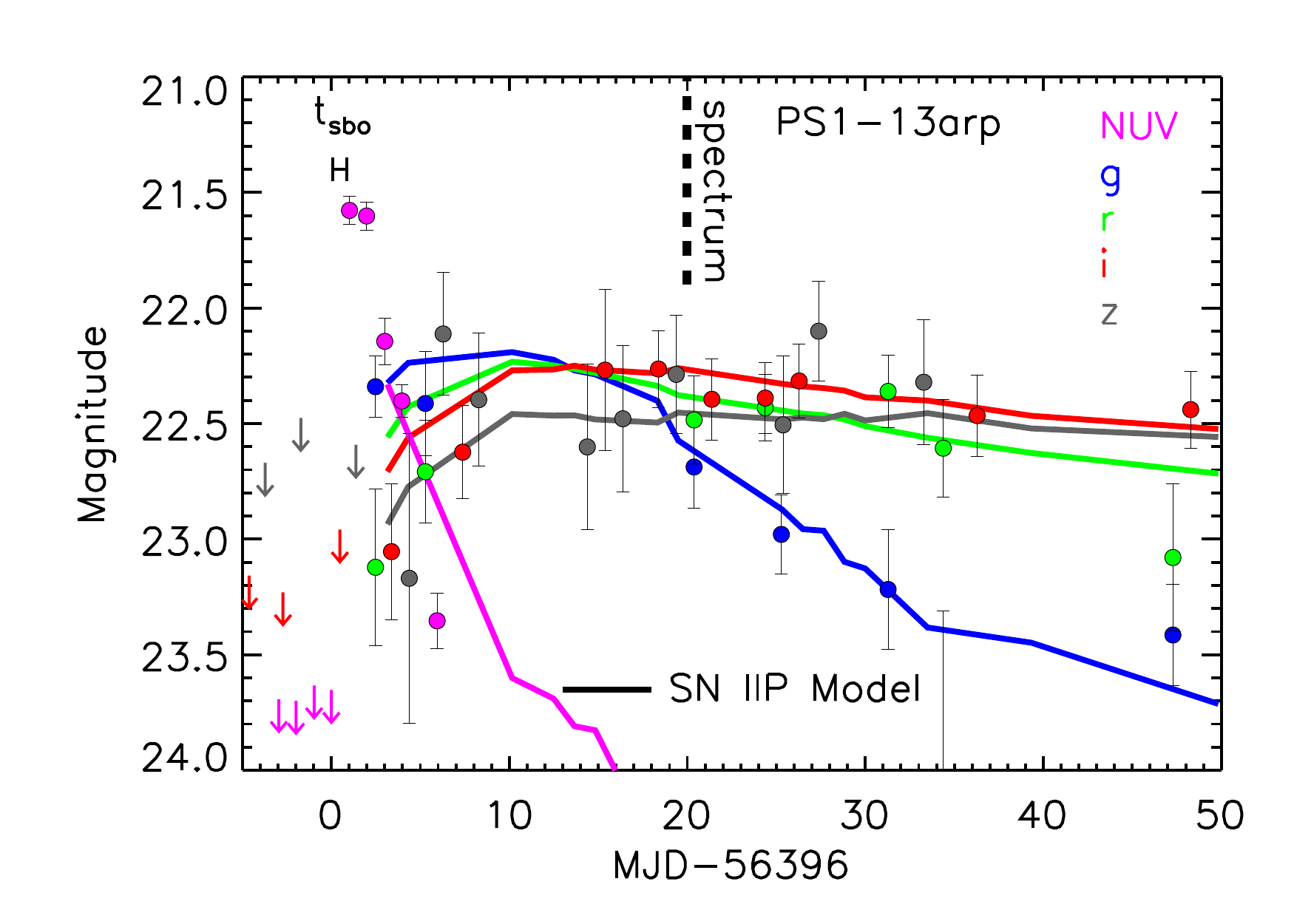}
%\begin{figure*}[htp]
%\begin{center}
%\includegraphics[scale=0.5]{toto_sn_500491_new_long_nomodel.eps}
\caption{\textsl{GALEX} $NUV$ and PS1 $\gps,\rps,\ips,\zps$\ light curve of PS1-13arp with 1$\sigma$ error bars, and arrows indicating 3$\sigma$ upper limits.  Dotted magenta line traces the $\sim 1$ day UV burst.  Thick black line indicates the time of the Gemini GMOS-N spectrum.  Thick solid lines show the CMFGEN model for SN 2006bp (modified to the rest frame of SNLS-04D2dc at $z=0.1854$, close in redshift to PS1-13arp) in the $NUV$, $g$, $r$, $i$, and $z$ bands from Dessart et al. (2008) scaled by a factor of $(1+z_{\rm 13arp})$ in the time axis to account for time dilation.  We use a minimized $\chi^2$ fit to the 4 optical bands in order to determine the time of shock breakout ($t_{\rm sbo}$) of MJD 56396.49 $\pm$ 0.40 in PS1-13arp, or 10 $\pm 8$ rest-frame hours after the last $NUV$ non-detection, and a vertical offset that is applied to all bands, of $-0.058 \pm 0.036$ mag. \label{fig:lc}}

%\end{center}
\end{figure*}

\section{Comparison to Type IIP SN\lowercase{e}} \label{sec:disc}

In Figure \ref{fig:lc} we plot the \textsl{GALEX} $NUV$ plus PS1 \gps, \rps, \ips, \zps light curve of PS1-13arp.  The global trends in the light curve are a $\sim 1$-day $NUV$ burst, a rapid \gps-band decline of $\sim 2.5$ mag/(100 day), and an over 50 day plateau in the \gps and \ips bands.  The optical light curve is characteristic of a normal Type IIP SN, and is well fitted with the spectral synthesis model for Type IIP SN 2006bp \citep{Dessart2008} K-corrected to a similar redshift to PS1-13arp (the model for SNLS-04D2dc at $z=0.1854$ from \citet{Gezari2008}) (shown in Figure \ref{fig:lc}).  The excellent agreement of the colors in the plateau imply a very small amount of internal extinction ($A_{\rm V} < 0.1$ mag).  We applied a scaling of $(1+z_{\rm 13arp})$ in the time axis to account for time dilation, and we used a minimized $\chi^2$ fit to the 4 optical bands determine a vertical offset factor $-0.058 \pm 0.036$ applied to all bands, as well as a horizontal shift, inferring a time of shock breakout for PS1-13arp of MJD $56396.49 \pm 0.40$.  This explosion time, determined entirely from the optical bands, is consistent with the timing of the $NUV$ burst, and occurs 10 $\pm$ 8 rest-frame hours after the last $NUV$ upper limit.  It is also in excellent agreement with the time of explosion determined from Bayesian parametric model fitting to the optical light curve in \citet{Sanders2014} of MJD $56396.6 \pm 0.5$.  The absolute magnitude of the $r$-band plateau at 50 days of $M_p,r = -17.54 \pm 0.12$ mag \citep{Sanders2014} is 0.2 mag fainter than SN 2006bp, and well within the range of normal Type IIP luminosities.

\begin{figure*}
\centering
\begin{minipage}{0.45\textwidth}
\centering
\includegraphics[scale=0.5]{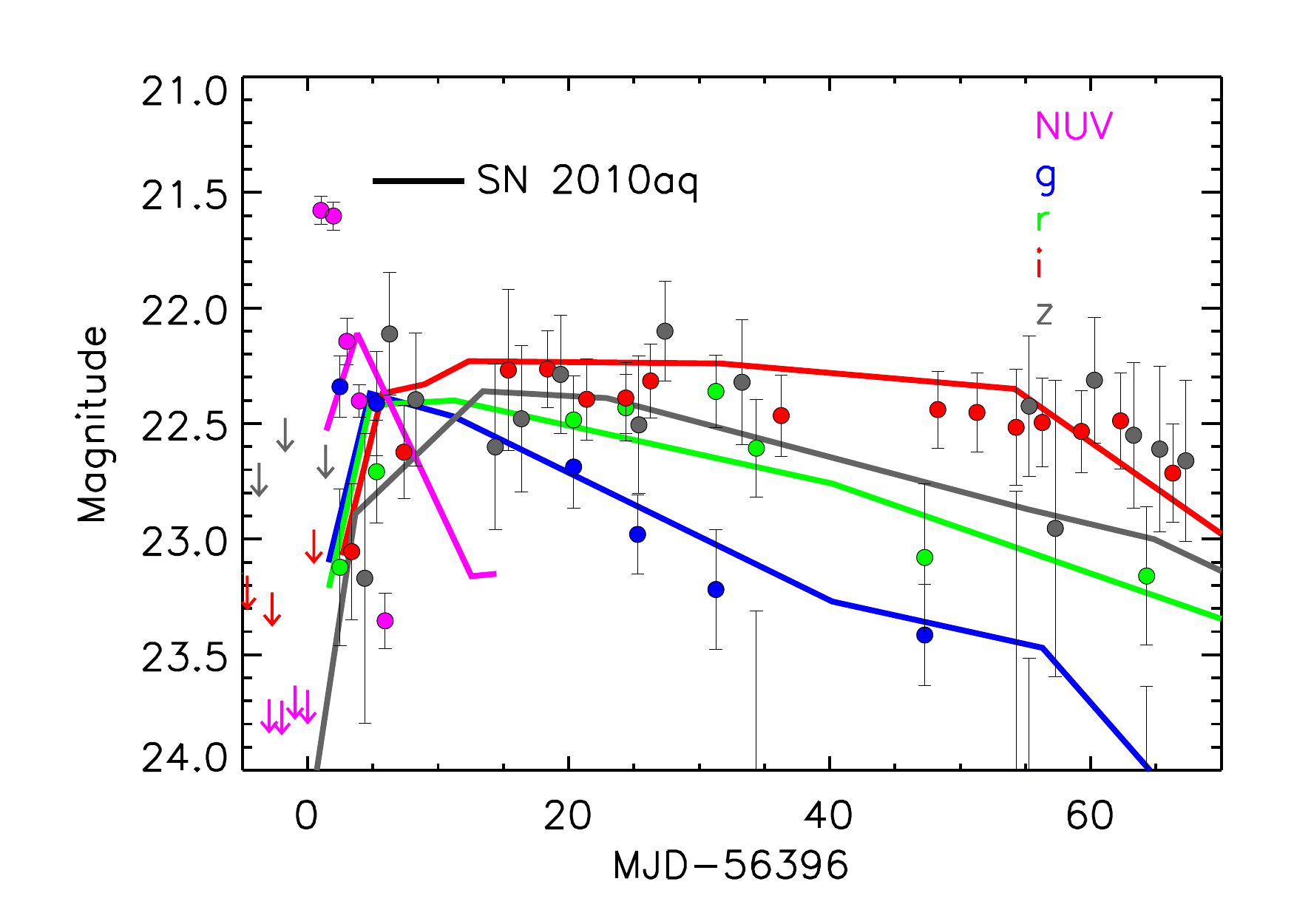} 
\end{minipage}%
\hspace{0.5cm}
\begin{minipage}{0.45\textwidth}
\centering
\includegraphics[scale=0.5]{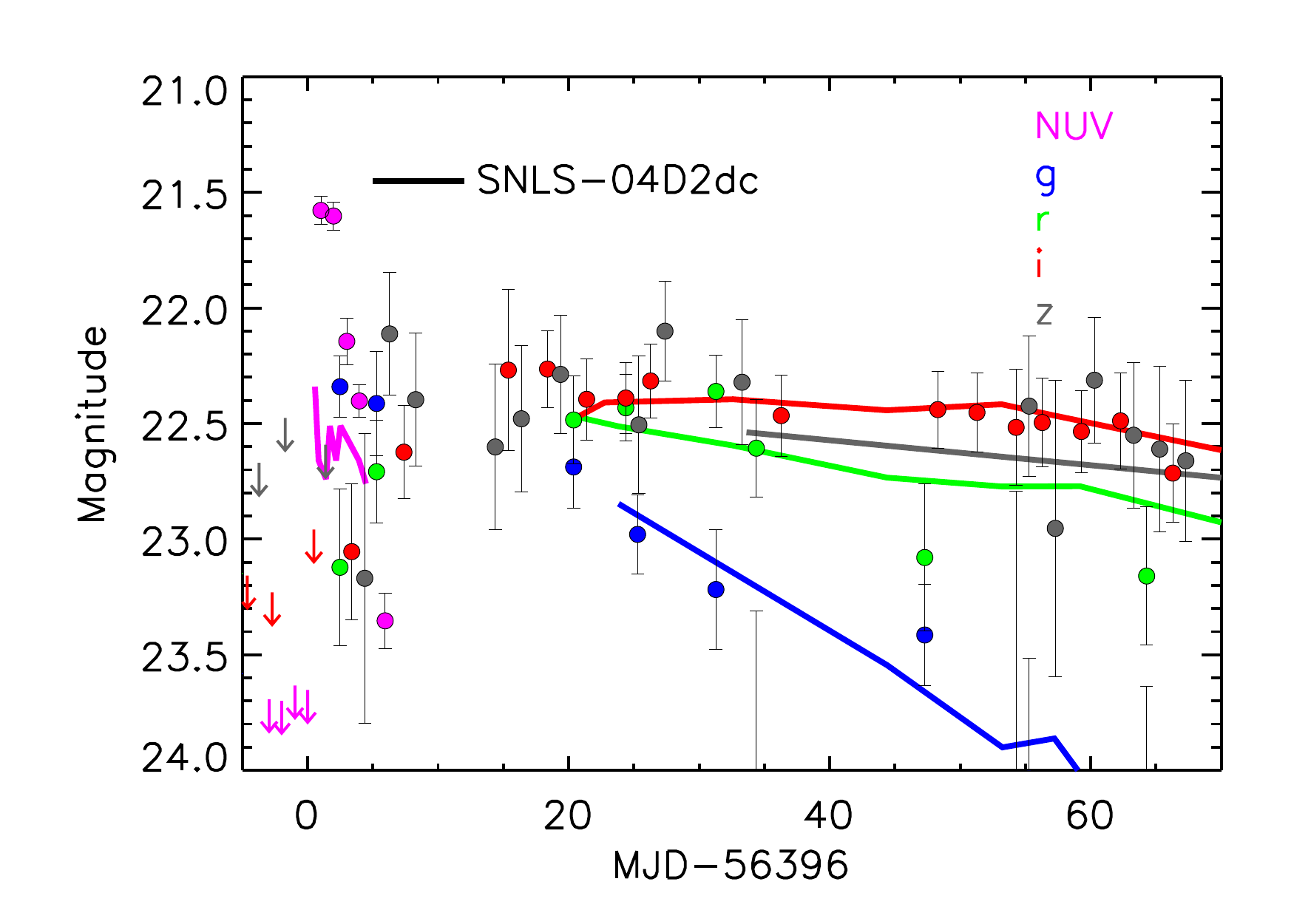}
\end{minipage}

\caption{Type II SNe with early UV photometry.  Comparison with SN 2010aq ({\it left}) and SNLS-04D2dc ({\it right}).  A constant offset of $+1.47$ mag has been applied to the SN 2010aq data from Gezari et al. (2010) and $-0.27$ mag to SNLS-04D2dc data from Gezari et al. (2008), plotted with solid thick lines.   For clarity, we do not plot the \textsl{GALEX} $FUV$ light curve of SNLS-04D2dc, which is lower signal-to-noise than the $NUV$ light curve.
\label{fig:comps1}
}
\end{figure*}

We calculate the peak absolute magnitude of the $NUV$ burst to be, 

\[ M_{\rm NUV} = m - DM - A_{\rm Gal} = -18.0~{\rm mag}, \]

\noindent where we use $m = 21.6$ mag, $d_{\rm L} = 800$ Mpc, $DM = 39.5$ mag, and Galactic extinction $A_{\rm Gal} = 8.2 E(B-V) = 0.13$ mag, where $E(B-V) = 0.016$ mag.  Note that for a hot blackbody temperature of $\approx 10^5$ K, the \textsl{GALEX} $NUV$ effective wavelength, $\lambda_{\rm rest} = 2316 \AA/(1+z) = 1985 \AA$ is on the Rayleigh-Jeans tail, and the K correction could be as large as $K(z) \approx -7.5\log(1+z) = -0.5$ mag.  The peak UV magnitude is consistent with the theoretical predictions for the peak of shock breakout in a RSG \citep{Gezari2008, Schawinski2008, Nakar2010, Rabinak2011}, and is brighter than the post shock breakout cooling peak that has definitively been observed in SNe IIP, SNLS-04D2ck with $NUV = -17.0$ mag \citep{Gezari2008} and SN 2010aq with $NUV = -17.3$ mag \citep{Gezari2010}. In Figure \ref{fig:comps1} we show a comparison of UV light curves from \textsl{GALEX} for two Type IIP SNe, caught rising in the UV, SN SNLS-04D2dc \citep{Gezari2008} and SN 2010aq \citep{Gezari2010}.  The comparison SNe have been offset in the vertical direction, and corrected for time-dilation.  While the optical behavior of the SNe is quite similar, there is a clear excess again in the UV burst of PS1-13arp, implying a different source of energy from the cooling envelope model fitted successfully to these comparison SNe. \\

\section{Comparison to Models} \label{sec:models}

\subsection{Shock Breakout in the Stellar Envelope} \label{sec:sbo}

While a burst of neutrinos marks the time of core-collapse, the supernova shock breakout is the first electromagnetic signature of a SN explosion.  The supernova shock ``breaks out''  when the radiative diffusion timescale, 

\[t_{\rm d} \sim \frac{3(\delta R)^2}{\lambda c} = \frac{3 \kappa \rho (\delta R)^2}{c} \approx \frac{3\tau H_\rho}{c}, \] 

\noindent where $\delta R$ is the depth of the shock, $\lambda = 1/\kappa \rho$ is the photon's mean free path, $\kappa$ is the opacity, $\tau$ is the optical depth, and $H_\rho$ is the atmospheric scale height, is shorter than the shock travel time, $t_s = \delta R/v_s$ \citep{Arnett1996}.  For a standard RSG envelope density and radial structure, the duration of the radiative precursor is  expected to be $\sim 1000$ sec \citep{Klein1978, Matzner1999}.  However, the shock breakout pulse will be prolonged up to $\sim 1$ hr due to light-travel time effects in the largest RSGs \citep{Levesque2005}, or up to days in the presence of a dense circumstellar wind \citep{Ofek2010, Chevalier2011, Moriya2011}.    

The rise time and duration of the UV burst in PS1-13arp are a factor of greater than $\sim$ 50 longer than both the radiative diffusion time at shock breakout from \citet{Matzner1999}:\\

\[ t_{\rm sbo} = 860 k^{-0.58} (\rho_1/\rho_*)^{-0.28} E_{51}^{-0.79} M_{15}^{0.21} R_{500}^{2.16}~{\rm sec},\]

\noindent where $R_{500}$ is the stellar radius in units of 500 $R_\odot$, $M_{10}$ is the stellar mass in units of 10 $M_\odot$, and $E_{51}$ is the energy of the explosion in units of $10^{51}$ erg s$^{-1}$,  and the light-crossing time of a RSG:\\

\[ t_{lc} = R_\star/c = 1160R_{500}~{\rm sec}. \]

\noindent Even assuming an envelope density a factor of 1000 less dense than the standard stellar structure models of \citet{Woosley2007}, as was done in \citet{Schawinski2008}, one can increase $t_{\rm sbo}$ only by a factor of $\sim$7.  However, such an extended, tenuous structure is better attributed to a circumstellar shell of material rather than the stellar envelope itself \citep{Falk1977}.  The long duration of the UV burst suggests that either the shock is "breaking out" in a wind around the progenitor star, or it is associated with the post-shock-breakout cooling envelope emission previously detected in the UV in Type IIP SNe with \textsl{GALEX} \citep{Gezari2008, Schawinski2008, Gezari2010, Ganot2014}.   In the following sections, we address both scenarios. \\

\subsection{Cooling Envelope Emission}

In the cooling envelope scenario, the \textsl{GALEX} observations did not catch the brief radiative precursor from shock breakout, but rather the UV burst is from the adiabatically expanding shock-heated SN ejecta.  As the hot envelope expands and cools, the peak energy of the thermal emission shifts from the soft X-rays into the UV when $3kT = h\nu_{NUV}(1+z)$, where $\nu_{NUV} = 2316$ \AA (or $T = 2.09$~eV), causing a peak in the $NUV$ light curve.  The rate of cooling depends sensitively on the radius of the progenitor, since the smaller the progenitor, the more energy is lost to adiabatic expansion.  Using the analytical model for the cooling envelope emission for a RSG, the largest likely Type II SN progenitor, from \citet{Nakar2010}, this occurs at a time :

\[t_{\rm peak, NUV} = 1.9 M^{-0.23}_{15}R^{0.68}_{500}E^{0.20}_{51} {\rm days}. \]

The rise-time in PS1-13arp is constrained from the time between the last $NUV$ non-detection and the first $NUV$ detection to be $< 0.87$ days in the SN rest-frame, or from the fitted time of shock breakout from \S \ref{sec:disc} to be $< 0.5 \pm 0.3$ d.  In order for this UV rise-time to be compatible with PS1-13arp, it would require $M^{-0.23}_{15}R^{0.68}_{500}E^{0.20}_{51} < 0.46$.  The UV peak can occur much sooner for a smaller star, such as a BSG ($t_{\rm peak} \approx  0.54$ days).  However, the late-time optical light curve of PS1-13arp is incompatible with such a compact progenitor.  For a BSG, with a typical radius of a factor of 10 smaller than a RSG, one would expect a redder and shorter-duration plateau \citep{Kleiser2011, Dessart2013}, which is inconsistent with the excellent fit of the plateau of SN PS-13arp to the SN 2006bp model.

\begin{figure}
\begin{center}
\includegraphics[scale=0.4]{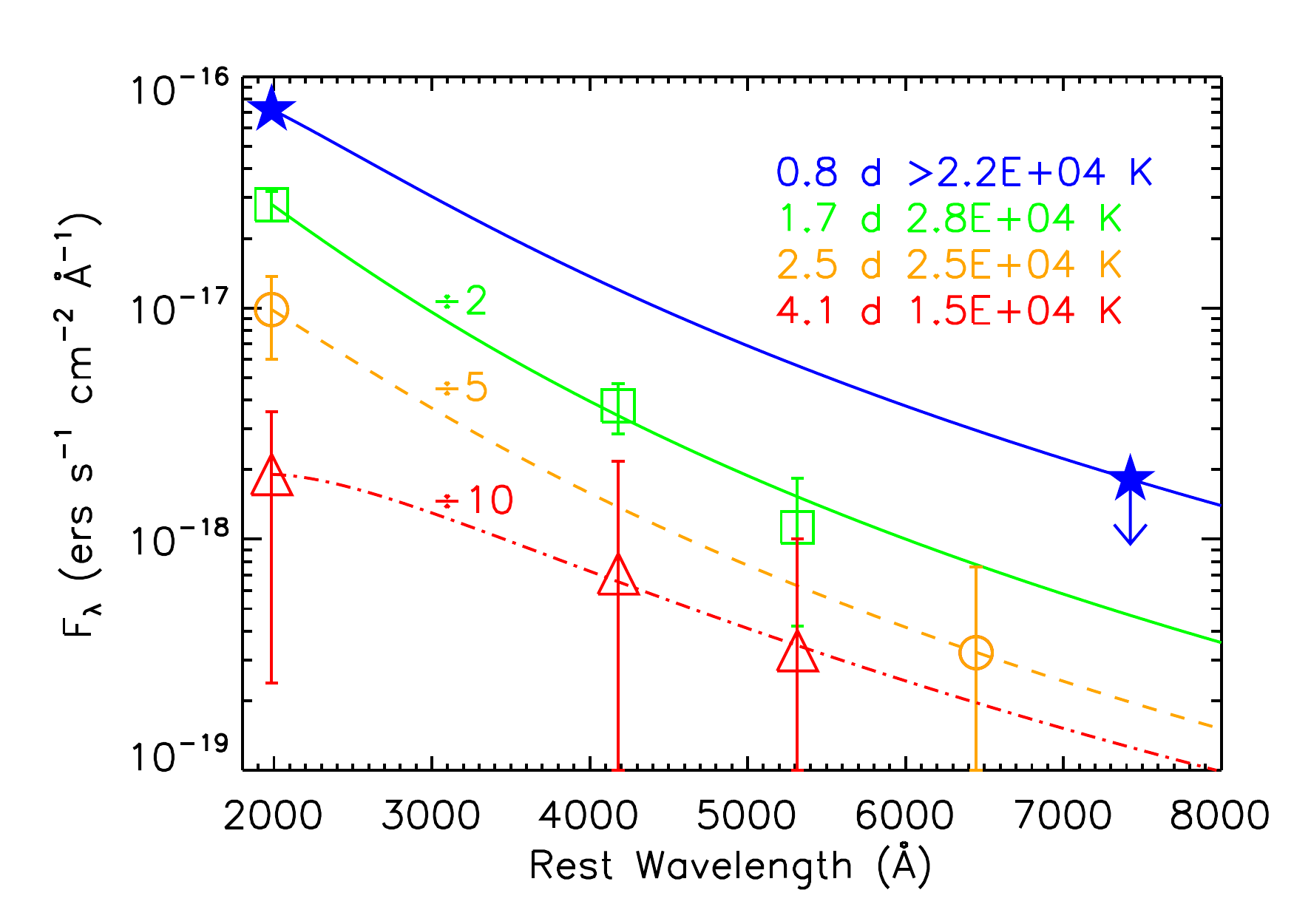}
\caption{Interpolated spectral energy distributions (SEDs) of PS1-13arp with time, and their least-squares fitted blackbody temperatures.  The time of each epoch is labeled in rest-frame days since the estimated time of shock breakout of MJD 56396.5 from Figure \ref{fig:lc}.  The temperature of the earliest epoch is constrained by a $z$-band upper limit to be $T > 2.2\times 10^{4}$ K.  The SEDs for 1.7 d, 2.5 d, and 4.1 d, have been divided by factors of 2, 5, and 10 respectively, for clarity.
\label{fig:sed}
}
\end{center}
\end{figure}

\begin{figure*}
\begin{center}
\includegraphics[scale=0.75]{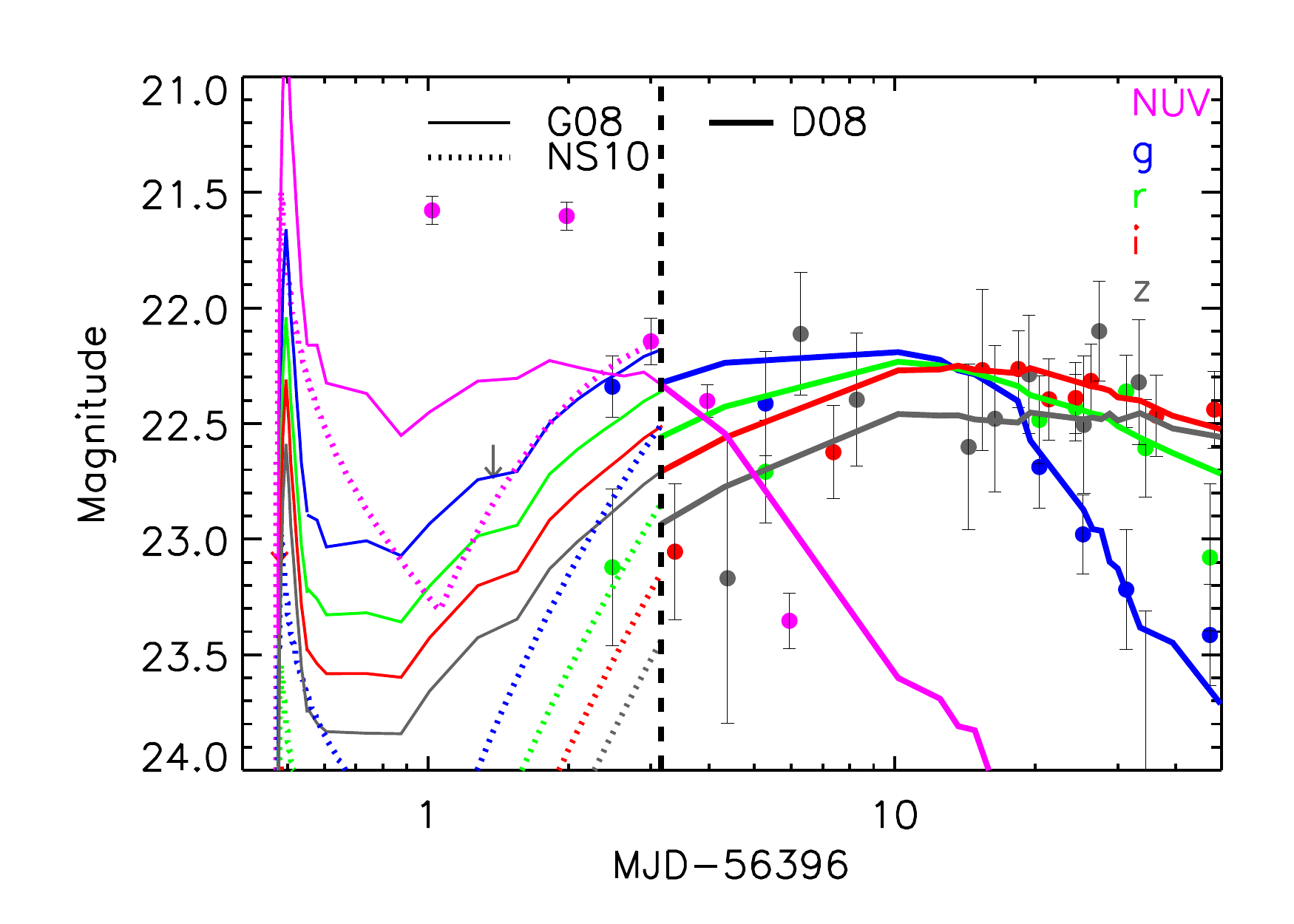}
\caption{Light curve of PS1-13arp on a logarithmic time scale, in order to compare the early UV emission to models for shock breakout and the cooling envelope in more detail.  Thick lines show the SN IIP model from \citet{Dessart2008} fitted to PS1-13arp in Figure \ref{fig:lc}, thin solid lines show the numerical model for shock breakout from the stellar surface and its cooling envelope from \citet{Gezari2008} (G08) for a RSG with $R=865 R_\odot$, $M = 12.58 M_\odot$, and $E = 1.2\times10^{51}$ erg, dotted lines show the analytical surface shock breakout and cooling envelope model from \citet{Nakar2010} (NS10) for $R = 500 R_\odot$, $M = 15 M_\odot$, and $E = 1\times10^{51}$erg.  The G08 model has been shifted vertically to overlap with the D08 model fit to PS1-13arp at $t > 3$ d in the $NUV$.  The NS10 model has been shifted vertically to match the $NUV$ data point for PS1-13Rp at $t\sim3$ d.  Note the clear excess in the observed $NUV$ emission above the cooling envelope emission components for both models at $t \sim 1-2$ d.  
\label{fig:model}
}
\end{center}
\end{figure*}

Figure \ref{fig:sed} shows the evolution of the UV/optical spectral energy distribution (SED) of PS1-13arp over time, with $NUV$ fluxes linearly interpolated to the time of the PS1 observations.  Unfortunately, we do not have a strong constraint on the temperature of the UV burst.  We can only put a lower limit from the $z$-band non-detection of $T > 2.2\times 10^{4}$ K.  Given that the next epoch, at a phase ($\phi$) of 1.7 rest-frame days since our estimated time of shock breakout, is hotter, with a least-squares fitted temperature of $T=(2.8 \pm  0.3) \times 10^{4}$ K, we adopt this as our lower-limit to the temperature of the UV burst at $\phi = 0.8$d, since the SN should be cooling with time.  The $NUV$ flux density, corrected for Galactic extinction, then corresponds to a peak bolometric luminosity of $L_{\rm bol} > 1.7 \times 10^{43}$ erg s$^{-1}$, which is an order of magnitude larger than expected for the cooling envelope at the time of the $NUV$ peak from \citet{Nakar2010}:  

\[ L_{\rm RSG, peak} = 2.70 \times 10^{42} M_{15}^{-0.87}R_{500}E_{51}^{0.96} {\rm erg s}^{-1}.\]

The observed $NUV$ absolute magnitude of PS1-13arp is also brighter than the corresponding peak absolute magnitude of this model of $M_{\rm NUV} \sim -16$ mag.  For a BSG, the luminosity at the time of the $NUV$ peak is even lower ($L_{\rm BSG, peak} \sim 8 \times 10^{41}$ erg s$^{-1}$).  Figure \ref{fig:model} compares the early UV/optical light curve of PS1-13arp with the RSG SN shock breakout and cooling envelope emission model from \citet{Nakar2010}, scaled to match the late-time SN 2006bp model fit.  While the observations and model are in good agreement close to the cooling envelope model peak and later ($t > 2$d), there is a clear prolonged UV excess above the model at earlier times.   We conclude that the peak of the observed UV burst is more luminous than expected for the cooling envelope emission following shock breakout, and thus we now investigate whether the UV burst could be from the shock breakout pulse itself, but prolonged in time due to the presence of a circumstellar wind.

\subsection{Shock Breakout in a Circumstellar Wind} \label{sec:wind}

The "shock breakout" is defined as when the radiation from the supernova shock is able to escape in front of the shock, i.e. when the radiative diffusion time is shorter than the shock travel time (see \S \ref{sec:sbo}).  For a bare star, this occurs in the outer envelope of the star (not actually at its surface).  However, when the star is enshrouded in a dense wind, the shock breakout is delayed, and occurs outside the surface of the star, in the circumstellar wind.  The theoretical literature still refers to this phenomenon as the shock breakout \citep{Chevalier2011, Balberg2011, Ofek2013}, even though the energy deposited from the shock is diffusing out (escaping) into the wind, instead of the stellar envelope.   A radiation-dominated shock in a uniform medium has a characteristic thickness in units of optical depth of $\tau_s \approx c/v_s$.  The shock will breakout in the wind if $\tau_w > \tau_s$, where $\tau_w$ is the thickness of the shock traveling through the wind in units of optical depth.  Here we assume the wind is from steady mass loss from the progenitor star $\dot{M}$, resulting in a density distribution $\rho_w = \dot{M}/4\pi r^2 v_w \equiv Dr^{-2}$, where $D \equiv 5.0 \times 10^{16} (\dot{M}_{0.01}/v_{w,10}) $~g cm$^{-1}$, and $\dot{M}_{0.01} = \dot{M}/(0.01 M_\odot$~yr$^{-1}$) and we assume the slow wind velocity characteristic of a RSG, $v_{w,10} = v_w/(10$~km s$^{-1}$).  \citet{Chevalier2011} approximate the optical depth for a shock wave traveling through the wind as $\tau_w \approx R_d^{-1}\kappa D$, where $R_d$ is the depth of the shock.  Hence shock breakout occurs when $\tau_w \approx c/v_{s}$, such that $R_d = \kappa D v_s/c$. The luminosity of the breakout radiation is given by $L_{bo} \approx E_{\rm rad}/t_{\rm bo}$, where $E_{\rm rad}$ is the radiative energy in the breakout shell, and $t_{\rm bo} = R_d/v_s = \kappa D/c = 6.6 k(\dot{M}_{0.01}/v_{w,10})$ days, where  $t_{\rm bo}$ is the time of shock breakout and $k = \kappa/(0.34$~cm$^{2}$~g$^{-1}$).  This is given in \citet{Chevalier2011} as $L_{\rm bo} \approx E_{\rm rad}/t_{\rm bo}$, and thus:\\

\[ L_{\rm bo}= 6.0 \times 10^{43} k^{-0.6} E_{51}^{1.2}M_{15}^{-0.6}(\dot{M}_{0.01}/v_{w,10})^{-0.2}~{\rm erg~s}^{-1}, \]

\noindent with a corresponding temperature of:\\

\[ T_{\rm rad} = 1.1 \times 10^5 k^{-0.25}(t_{d,1})^{-0.25}~K, \]

%\citet{Ofek2010} estimate the luminosity from the rate that the shock kinetic energy is converted to radiation: \\

%\[ L_{\rm bo} \approx 4\pi r_s^2 v_s \left (\frac{1}{2}\rho_s v_s^2\right ) = 1.0 \times 10^{43} (\dot{M}_{0.01}/v_{w,10}) (E_{51}/M_{10})^{3/2}~{\rm erg~s}^{-1}\]

%\noindent and a temperature of:\\

%\[ T_{\rm rad} = 1.5 \times 10^5 k^{-0.25}(t_d/ 1~{\rm day)^{-0.25}~K} \]

\noindent where $t_{d,1} = t_d/(1~{\rm day}$).  Given the observed rise-time of $t_{\rm rise} \aplt 0.87$ d in the SN rest-frame, and the weak dependence of $T$ on $t_{d}$, we assume a shock breakout radiation temperature of $T \approx 10^5$ K, which is allowed by our observed SED at the UV peak  based on the $z$-band upper limit during the UV burst.  Given our observed Galactic-extinction corrected luminosity density in the $NUV$ of $L_{\nu_e=(1+z) \nu_o} = f_{\nu_o} 4\pi d_{\rm L}^2/(1+z) = 6.25 \times 10^{27}$ erg s$^{-1}$ Hz$^{-1}$, and assuming the radiation from an optically thick blackbody, for which $L_\nu = (2\pi h/c^2)\nu^3/(e^{h\nu/kT} - 1)4\pi R_{\rm bb}^2$, one gets $R_{\rm bb} = 5.8 \times 10^{13}$ cm and a bolometric luminosity of $2.4 \times 10^{44}$ erg s$^{-1}$ (a factor of 14 greater than the observed lower limit).   The inferred radius is close to the upper range of radii measured for RSG ($2-6 \times 10^{13}$ cm; \citet{Levesque2005}), and potentially situated {\it outside} the stellar envelope, in the circumstellar wind.  Given the rise-time of $\aplt 0.87$ day, this radius then corresponds to a shock velocity of $v_s \apgt 7 \times 10^{3}$ km s$^{-1}$, similar to the velocity width of the detected broad H$\alpha$ line.

\begin{figure*}
\begin{center}
\includegraphics[scale=0.75]{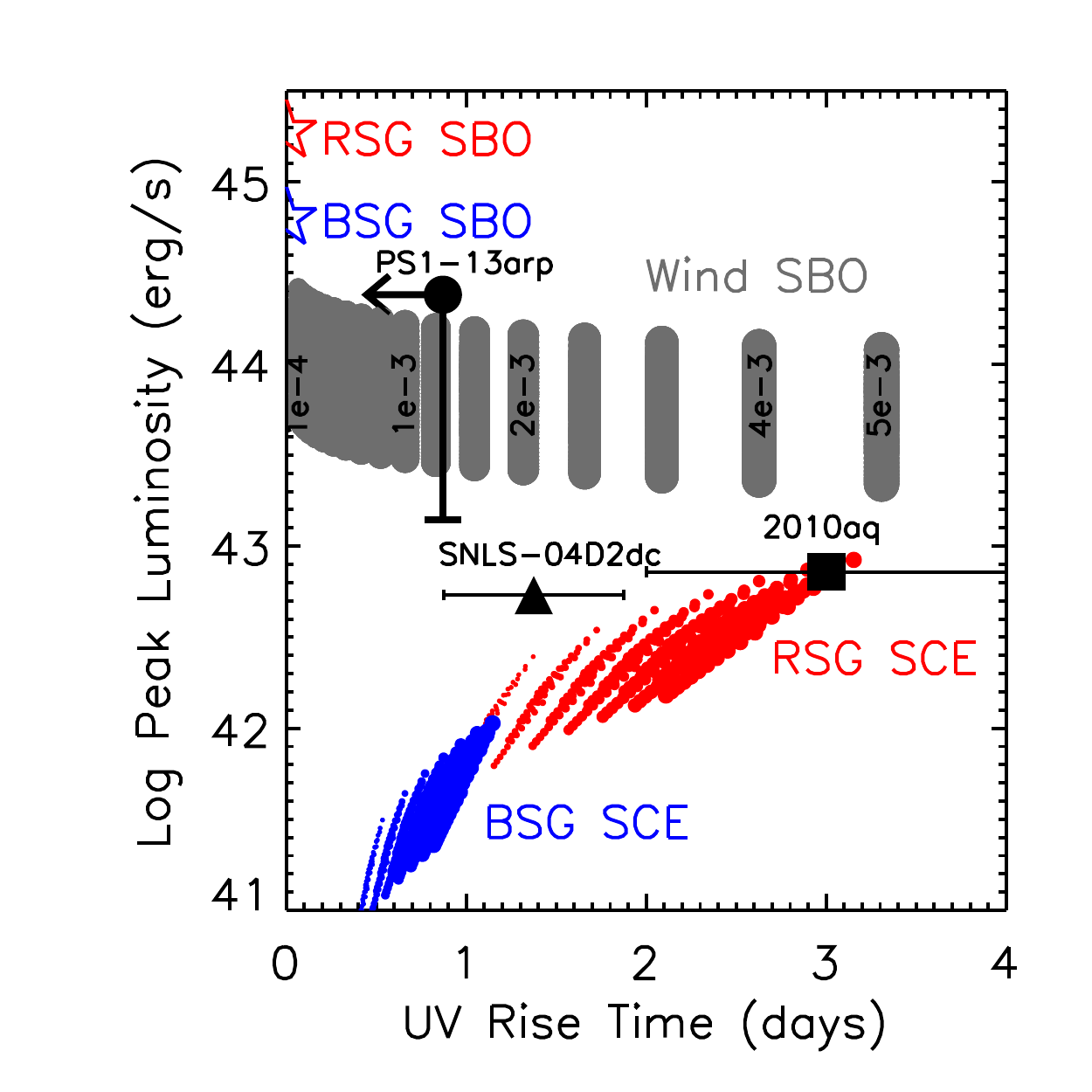}
\caption{Parameter space of peak bolometric luminosity vs. the rise time to peak in the $NUV$ covered by models for shock breakout in the stellar envelope for a RSG (red star) and BSG (blue star), shock breakout into a wind (grey bars), and shock cooling envelope for a BSG (blue dots) and RSG (red dots) with a range of values for $\dot{M}$, $R$, $M$, and $E$.  The size of the dots are scaled by the relative size of $R$ within the range of values of $250-850 R_\odot$ for RSGs, and $25-85 R_\odot$ for BSGs.  The ranges of the other SN parameters are $M=10-25 M_\odot$, $E = (0.5-1.5)\times10^{51}$ erg.  The wind mass-loss rate ranges from $\dot{M} = (.01-5) \times10^{-3} M_\odot$/yr, with some values labeled on the diagram.  The peak bolometric luminosity, and UV rise time are plotted for PS1-13arp (black dot), SN 2010aq (black square), and SN SNLS-04D2dc (black triangle).  The peak bolometric luminosity of PS1-13arp is plotted ranging from a lower limit for the observed SED with $T > 2.8\times10^{4}$ K, up to the expected wind shock breakout model temperature of $T\approx10^5$ K.
\label{fig:models}
}
\end{center}
\end{figure*}

The \citet{Chevalier2011} model places two independent constraints on the mass-loss rate ($\dot{M}$).  The first is from the timescale of the event.  The observed rise time corresponds to $\dot{M} \aplt 1.3\times10^{-3} (v_{w,10}/k) t_{d,1}~M_\odot$ yr$^{-1}$.  The duration of the shock breakout signal in the \citet{Chevalier2011} wind model is on the order of the rise-time to peak.  Thus the rise time in PS1-13arp cannot be much shorter than the duration of the UV burst of $\apgt$ 0.82 d in the SN rest frame.  Thus we favor a mass-loss rate closer to $10^{-3} M_\odot$ yr$^{-1}$.  The observed luminosity corresponds to $\dot{M} \sim 1\times10^{-5} (v_{w,10}) k^{-3} E_{51}^6 M_{15}^{-3}~M_\odot$ yr$^{-1}$.    Given the strong sensitivity of the peak luminosity to $E_{51}$, this could be easily reconciled with the mass-loss rate estimated from the timing of the event by adopting an explosion energy of $E_{51} \sim 1.5$.  Figure \ref{fig:models} plots the parameter space for peak bolometric luminosity versus rise time to peak in the $NUV$ for shock breakout in a wind, and cooling envelope emission for BSGs and RSGs with a range of $E_{51} = (0.5-1.5)\times10^{51}$ erg, $R = 250-850 R_\odot$ for the RSG and $R = 25-85$ for the BSG, and $M = 10-25 M_\odot$.  The observed peak luminosity of PS1-13arp is well above that expected for the cooling envelope phase, and is within the range of luminosities and timescales for a burst powered by shock breakout in a wind.   

The upper limit on $\dot{M}$ from the observed rise time is consistent with the lack of excess thermal radiation in the optical bands at later times.  \citet{Moriya2011} model the light curves of Type IIP SNe as a function of $\dot{M}$, and find that for $\dot{M} > 10^{-3} M_\odot$~yr$^{-1}$, there is an observable excess in the UV and optical bands due to the conversion of the kinetic energy of the ejecta into thermal energy via interaction with the CSM.   The mass-loss rates in RSG inferred from spectroscopic studies range from $10^{-7}-10^{-4} \dot{M}$ yr$^{-1}$ \citep{Mauron2011, vanLoon2005}. Thus while the inferred mass-loss rate from the luminosity and duration of the shock breakout is on the high end of the observed distribution, these studies are sensitive to the time-averaged mass-loss rate, and do not sample short-lived, episodic enhanced mass-loss that may occur in pre supernova-RSGs \citep{Meynet2014}, perhaps as a result of pulsation-driven superwinds \citep{Yoon2010}.  Unfortunately, we do not have sufficient signal-to-noise in the spectrum of PS1-13arp to look for a high-velocity absorption feature or "notch", predicted to be a signature of the excitation of ejecta by X-rays produced by circumstellar interaction \citep{Chugai2007}.

X-ray and radio observations of Type IIP SNe also put conservative upper limits on the mass-loss rates of pre-SN RSGs to $< 10^{-6}-10^{-5} M_\odot$ yr$^{-1}$ from the lack of thermal bremsstrahlung emission or synchrotron radiation from the interaction of the SN shock wave with the CSM \citep{Dwarkadas2014, Chevalier2006}.  Radio and X-ray emission from Type IIP SN 1999em was interpreted as the result of circumstellar interaction with a stellar wind of $\dot{M} \sim 2\times 10^{-6} M_\odot$ yr$^{-1}$ \citep{Pooley2002}.  Mass-loss rates should scale with RSG luminosity, and thus with stellar mass.  \citet{Sanders2014} were not able to put a direct constraint on the progenitor mass of PS1-13arp from the comparison of the bolometric light curve to hydrodynamic models, however they did infer a ejected nickel mass from a comparison of the late-time bolometric light curve of PS1-13arp to SN 1987A of $M_{\rm Ni} = 0.34 ^{+0.20}_{-0.22} M_{\odot}$.  This amount of nickel is on the high-mass tail of the distribution for their SNe IIP sample, and could be indicative of a more massive progenitor RSG for PS1-13arp.   Another clue is the weak P-Cygni absorption in the H$\alpha$ profile of PS1-13arp at $t \sim 17$ d.  Weak P-Cygni absorption has been found in brighter than normal Type IIP SNe, and explained by lower envelope masses and higher mass-loss rates \citep{Gutierrez2014}.

\section{Conclusions} \label{sec:conc}

We present the joint \textsl{GALEX} UV and PS1 optical discovery of a Type IIP SN with a normal optical light curve, including the trademark plateau, preceded by a UV burst with the peak luminosity expected from shock breakout, but with a longer than expected duration.  We rule out shock cooling emission from a compact BSG progenitor, and favor a shock breakout from a RSG with a high, but not extreme, amount of pre-SN mass-loss ($\approx 10^{-3} M_\odot$ yr$^{-1}$).  This mass-loss rate is high enough to increase the height at which the SN shock breaks out to beyond the stellar surface, but not large enough to effect the optical light curve or spectrum measured weeks after the explosion.  In particular, we do not detect narrow emission lines that could be powered by interaction with a dense CSM (from $\dot{M} \sim 0.01 M_\odot$ yr$^{-1}$ to $0.1 M_\odot$ yr$^{-1}$) as seen in Type IIn SNe \citep{Turatto1993, Chugai1994, Smith2008}

However, we must caution that the theoretical models discussed here have yet to be verified in the normal case of shock breakout in the stellar envelope.  In the case of Type IIP SN SNLS-04D2dc, the early \textsl{GALEX} UV photometry could be fitted either with a delayed shock breakout in a non-standard, extended, low-density envelope of a RSG followed by the cooling envelope \citep{Schawinski2008}, or a numerical simulation for only the cooling envelope component \citep{Gezari2008}.  The cadence ($\sim 1.5$ hr) and low signal-to-noise photometry (S/N $\sim 2-4$) in this source were not enough to favor one model over the other, let alone constrain their detailed parameters.   While a wind effectively smears out the shock breakout signal, and makes it possible to resolve with a coarser (1 day) cadence, as we report here for PS1-13arp, this interpretation would have stronger footing if the theory were proven in the more common case of shock breakout in the SN envelope.  

Rapid-cadence observations on the timescale of minutes, particularly in the UV band, are required to resolve the putative rapid rise from shock breakout in the SN light curve, and its second UV peak due to the cooling envelope.  
Future observations of shock breakout of Type IIP SNe in the UV could be a sensitive probe of the mass-loss history of pre-SN RSGs, which is difficult to constrain directly \citep{Meynet2014}.  Furthermore, the presence of a wind will greatly enhance the ability of future UV time domain surveys, such as ULTRASAT \citep{Sagiv2014}, to directly detect this diagnostic phase of the SN explosion.

\acknowledgments

S. G. thanks the Gemini Deputy Director for approving a change of program for TOO program GN-2013A-Q-32 to observe PS1-13arp.  The Pan-STARRS1 Surveys (PS1) have been made possible through contributions of the Institute for Astronomy, the University of Hawaii, the Pan-STARRS Project Office, the Max-Planck Society and its participating institutes, the Max Planck Institute for Astronomy, Heidelberg and the Max Planck Institute for Extraterrestrial Physics, Garching, The Johns Hopkins University, Durham University, the University of Edinburgh, Queen's University Belfast, the Harvard-Smithsonian Center for Astrophysics, the Las Cumbres Observatory Global Telescope Network Incorporated, the National Central University of Taiwan, the Space Telescope Science Institute, the National Aeronautics and Space Administration under Grant No. NNX08AR22G issued through the Planetary Science Division of the NASA Science Mission Directorate, the National Science Foundation under Grant No. AST-1238877, the University of Maryland, and Eotvos Lorand University (ELTE).

\newpage

\bibliographystyle{fapj}

\newpage

\end{document}